\documentclass[conference]{IEEEtran}
\IEEEoverridecommandlockouts
\usepackage{cite}
\usepackage{amsmath,amssymb,amsfonts}
\usepackage{algorithmic}
\usepackage{graphicx}
\usepackage{textcomp}
\usepackage[linesnumbered,ruled,vlined]{algorithm2e}
\usepackage{booktabs}
\usepackage{multirow}
\usepackage{geometry}
\usepackage{makecell}
\usepackage{subcaption}
\usepackage{multirow}
\usepackage[table,xcdraw]{xcolor}
\usepackage{svg}
\usepackage{url}

\def\BibTeX{{\rm B\kern-.05em{\sc i\kern-.025em b}\kern-.08em
    T\kern-.1667em\lower.7ex\hbox{E}\kern-.125emX}}
\begin{document}

\title{Towards a Decentralised Application-Centric Orchestration Framework in the Cloud-Edge Continuum
}
\author{\IEEEauthorblockN{Amjad Ullah\textsuperscript{\S}, Andras Markus\textsuperscript{†}
Hacı İsmail Aslan\textsuperscript{*}, 
Tamas Kiss\textsuperscript{\#}, \\
Jozsef Kovacs\textsuperscript{+},
James Deslauriers\textsuperscript{\#},
Amy L. Murphy\textsuperscript{‡}, 
Yiming Wang\textsuperscript{‡}, 
Odej Kao\textsuperscript{*}
} \\
\IEEEauthorblockA{
\textsuperscript{\S}A.Ullah@napier.ac.uk, Edinburgh Napier University, Edinburgh, United Kingdom \\
\textsuperscript{†}andras.markus@frontendart.com, FrontEndART Software Ltd., Szeged, Hungary \\
\textsuperscript{*}\{aslan, odej.kao\}@tu-berlin.de Technische Universität, Berlin, Germany \\
\textsuperscript{\#}\{T.Kiss, J.Deslauriers\}@westminster.ac.uk, University of Westminster, London, United Kingdom \\
\textsuperscript{+}jozsef.kovacs@sztaki.hu, Institute for Computer Science and Control (SZTAKI), Budapest, Hungary \\  
\textsuperscript{‡}\{murphy, ywang\}@fbk.eu, Fondazione Bruno Kessler, Trento, Italy}
}

\maketitle
\begin{abstract}
%In cloud-to-edge computing environments, selecting the optimal resources to meet the requirements of a distributed application is a critical challenge. This task, known as resource allocation or matchmaking, is a multi-objective optimization problem and involves optimizing multiple attributes such as price, latency, bandwidth, and energy consumption. In this paper, we first address the problem using a centralized approach, employing traditional optimization techniques such as weighted sum methods, attribute normalization, and multi-objective ranking strategies like the Borda method to score and rank available resources. Subsequently, we extend this solution into a decentralised framework, where resource agents operating as gateway peers in a simulated network collaboratively perform the allocation. By simulating the decentralised architecture in a network simulator, we assess how distributed decision-making impacts the performance of the system compared to the centralized method. The results provide valuable insights into the trade-offs between centralized and decentralised resource allocation approaches for cloud-to-edge applications, offering guidance for the design of scalable, efficient infrastructures.

The efficient management of complex distributed applications in the Cloud-Edge continuum, including their deployment on heterogeneous computing resources and run-time operations, presents significant challenges. Resource management solutions---also called orchestrators---play a pivotal role by automating and managing tasks such as resource discovery, optimisation, application deployment, and lifecycle management, whilst ensuring the desired system performance. This paper introduces Swarmchestrate, a decentralised, application-centric orchestration framework inspired by the self-organising principles of Swarms. Swarmchestrate addresses the end-to-end management of distributed applications, from submission to optimal resource allocation across cloud and edge providers, as well as dynamic reconfiguration. Our initial findings include the implementation of the application deployment phase within a Cloud-Edge simulation environment, demonstrating the potential of Swarmchestrate. The results offer valuable insight into the coordination of resource offerings between various providers and optimised resource allocation, providing a foundation for designing scalable and efficient infrastructures.

\end{abstract}


\section{Introduction} \label{sec:intro}
The Cloud-Edge continuum represents a dynamic ecosystem where computational resources are distributed across multiple administrative domains, ranging from centralised data centres to localised edge nodes, each with unique computational capabilities, constraints, and performance characteristics~\cite{moreschini2022cloud}. The rapid increase in such paradigms has transformed how modern distributed applications provision, manage and utilise computational resources. Orchestration solutions are typically employed to deal with the associated challenges of managing complex distributed applications in the Cloud-Edge compute continuum, including their deployment on heterogeneous computing resources and run-time operations ~\cite{svorobej2020orchestration, swarmchestrate, hu2023architectural, ullah2021micado}. 

The challenges include ensuring seamless and simultaneous access to the heterogeneous and decentralised resource landscape, along with effective coordination across diverse cloud, fog, and edge providers to enable end-to-end services. Additionally, achieving optimisation of multiple, often conflicting quality of service (QoS) goals---such as optimal placement, resource usage efficiency, application performance, cost, energy consumption, and security---remains a critical focus. Scalability of the continuum, adaptability to changing application and infrastructure conditions, and efficient monitoring to collect workload and resource usage statistics throughout the spectrum are also essential requirements.

%\begin{enumerate}
%    \item Seamless and simultaneous access to the heterogeneous and decentralised resource landscape of the Cloud-Edge compute continuum.
%    \item Coordination across diverse cloud, fog and edge providers to facilitate end-to-end services.
%    \item Simultaneously optimisation of multiple, often conflicting quality of service (QoS) goals specified in terms of optimal placement, optimising resource usage, application performance, cost, energy, security, etc.
%    \item Scalability of the compute continuum. 
%    \item Maintaining system adaptability to changing application and infrastructure conditions.
%    \item Efficient monitoring to collect the status of the workload and resource usage statistics across the entire spectrum of the continuum.
%\end{enumerate}
Addressing these challenges has drawn significant attention from industry and academia toward developing orchestration solutions for Cloud-Edge continuum~\cite{ullah2023orchestration, bohm2022cloud, bohm2022towards}. These solutions can generally be classified into centralised and decentralised approaches based on their control topology. Centralised approaches involve a single entity---typically run in the cloud---collecting data from the entire continuum to handle orchestration functions like resource selection, application deployment, monitoring, and execution control. This approach benefits from consistent decision-making and ease of implementation, however, suffers from significant drawbacks, including limited scalability, a single point of failure, and vulnerability to cyber-attacks~\cite{pradeep2021holistic}. Additionally, although well-suited to cloud-only systems, it raises critical concerns for the distributed continuum such as continuous data transfers from diverse administrative domains, privacy and security risks, and unreliable connectivity.

The decentralised approach, on the other hand, consists of multiple decision-making entities (orchestrators) distributed across the continuum, each operating independently for specific applications or within local resource domains. These orchestrators collaborate as needed to achieve global SLA goals. Several studies (see Section~\ref{sec:relatedwork}) have explored such strategies, demonstrating their potential to manage the complexities of the compute continuum while ensuring efficiency and SLA compliance. In the same realm, this paper introduces Swarmchestrate, inspired by the self-organising principles of Swarms. Swarmchestrate adopts an application-centric approach, advancing the concept of decentralised orchestration to prioritise application-specific goals, in contrast to existing solutions that primarily focus on optimising resource-provider objectives. More specifically, our key contributions include the following:
\begin{enumerate}
    \item A novel decentralised orchestration architecture based on the high-level concept defined in~\cite{swarmchestrate}.
    \item A simulation-based implementation demonstrating the potential of the proposed architecture.
    \item Implementation of a resource offer collection strategy from diverse resource providers.
    \item Implementation of several optimisation algorithms for resource selection.
    \item A thorough evaluation of the proposed approach from an application deployment viewpoint. 
\end{enumerate}
The remainder of this paper is structured as follows: Section~\ref{sec:relatedwork} provides a review of existing decentralised orchestration solutions. Section~\ref{sec:swarmchestrate} introduces the Swarmchestrate framework, detailing its architecture and key aspects. Section~\ref{sec:evaluation} outlines the experimental setup, methodologies and experimentation for Swarmchestrate's performance evaluation. Finally, Section~\ref{sec:conclusion} concludes the paper.

\section{Related Work} \label{sec:relatedwork}
Early orchestration research primarily explored centralised approaches, with notable initiatives like~\cite{kumara2021sodalite, ullah2021micado, verginadis2021prestocloud} and industry efforts such as HPE\footnote{HPE GreenLake: ke. https://www.hpe.com/us/en/gree}, Intel\footnote{ Intel Smart Edge Open: https://smart-edge-open}, Azion\footnote{Azion: https://www.azion.com/en/products/edge-orchestrator/}. However, the limitations of centralised methods have driven a shift towards decentralised orchestration, which is the focus of this section.

%Early research on orchestration focused mainly on centralised approaches. Some notable research initiatives include SODALITE@RT~\cite{kumara2021sodalite}, MiCADO-Edge~\cite{ullah2021micado}, PrEstoCloud~\cite{verginadis2021prestocloud}, Pledger~\cite{pledger-project} and the industry has also made significant contributions with systems such as~HPE GreenLake~\cite{hpe-greenlake}, Intel Smart Edge Open~\cite{intel}, Azion~\cite{azion}, ONAP~\cite{onap}. While centralised approaches offer simplicity and consistency in decision-making, they often struggle with scalability and become bottlenecks when managing resources across diverse infrastructures (e.g., cloud, fog, and edge). These limitations have shifted the interest towards decentralised orchestration, which is the focus of the remainder of this section.

Hierarchical or layered architectures have been proposed in various studies. For example, mF2C~\cite{masip2021managing} is an N-layered architecture for resource utilisation, from edge (Layer-N) to cloud (Layer-0). In this approach, mF2C agents are deployed at each layer to collaborate on service execution requests, prioritising the lowest layer to minimise latency. Oakestra~\cite{bartolomeo2023oakestra} employs a two-layered architecture with a root orchestrator and cluster orchestrators, enabling multiple edge and cloud providers to contribute their local deployments as independent clusters under separate administrative control. Cluster orchestrators manage resources locally, while the root orchestrator oversees the global infrastructure.

Fernandez et al.~\cite{fernandez2019enabling} propose an approach based on network slices, representing partitions of the IoT ecosystem tailored for specific customers. This approach uses a slice orchestrator that collaborates with multiple domain-specific cloud/edge resource orchestrators. The slice orchestrator manages slice creation and administration, while the domain orchestrators handle resource selection and deployment within their respective domains. A similar approach (i.e. network slices-based) is also adopted in~\cite{tusa2023end}. This approach, however, also introduced a marketplace-based mechanism where resource providers can offer their resources for inclusion in slices.

Several researchers followed a peer-to-peer-based model. For example, HYDRA~\cite{jimenez2020hydra} establishes a P2P overlay network where each computational node acts as a resource and an orchestrator, enabling the execution of application microservices. An orchestrator can then take on various roles: managing an entire application, overseeing a subset of services, or handling a single service. Caravela~\cite{pires2021distributed} used a similar approach. However, unlike HYDRA, it introduces a market-oriented approach, where volunteer resources join the ecosystem in exchange for compensation. Both of these approaches focused on scalability and robustness through dynamic partitioning, where nodes operate autonomously, with a distributed hash table and lookup algorithm, enabling discovery and communication within the network. 

Castellano et al.~\cite{castellano2019service} followed a dedicated orchestrator on a per application-based approach in contrast to the individual resource as orchestrator-based approach in~\cite{jimenez2020hydra,pires2021distributed}. Zeinab et al.~\cite{nezami2021decentralized} introduced EPOS Fog, a multi-agent system where each node is an agent that defines which service is deployed on which host in the agent's neighbourhood. Zolton's~\cite{mann2022decentralized} approach partitions the infrastructure into isolated segments called fog colonies, allowing independent optimisation within each. They evaluated decentralisation and coordination strategies, including optimisation limited to a single colony, coordinated optimisation across multiple colonies with isolated resource pools, and overlapping optimisation, where the same resource could belong to multiple colonies.

While previous research has made significant contributions, our approach is distinguished for its highly decentralised, application-centric focus and emphasis on self-organisation, enabling a more adaptive and resilient orchestration framework. More specifically,
unlike the purely hierarchical or P2P models in prior studies, our approach adopts a hybrid model. The hierarchical aspect stems from its two-layered structure: the interface---a dynamic network of distributed resource agents---contrasts with the static mF2C edge-layer agents~\cite{masip2021managing} and the root orchestrator in~\cite{bartolomeo2023oakestra}; and application spaces composed of Swarm agent networks managing individual applications. The operational mechanisms at each layer are designed around a P2P model. Additionally, unlike in existing approaches such as that in~\cite {jimenez2020hydra} and~\cite{pires2021distributed}, resources take on a passive role, being discovered by the interface layer based on the specific requirements of the applications.
\section{Swarmchestrate: Proposed approach}\label{sec:swarmchestrate}
\begin{figure*}[t]
    \centering
        \includegraphics[width=0.9\textwidth]{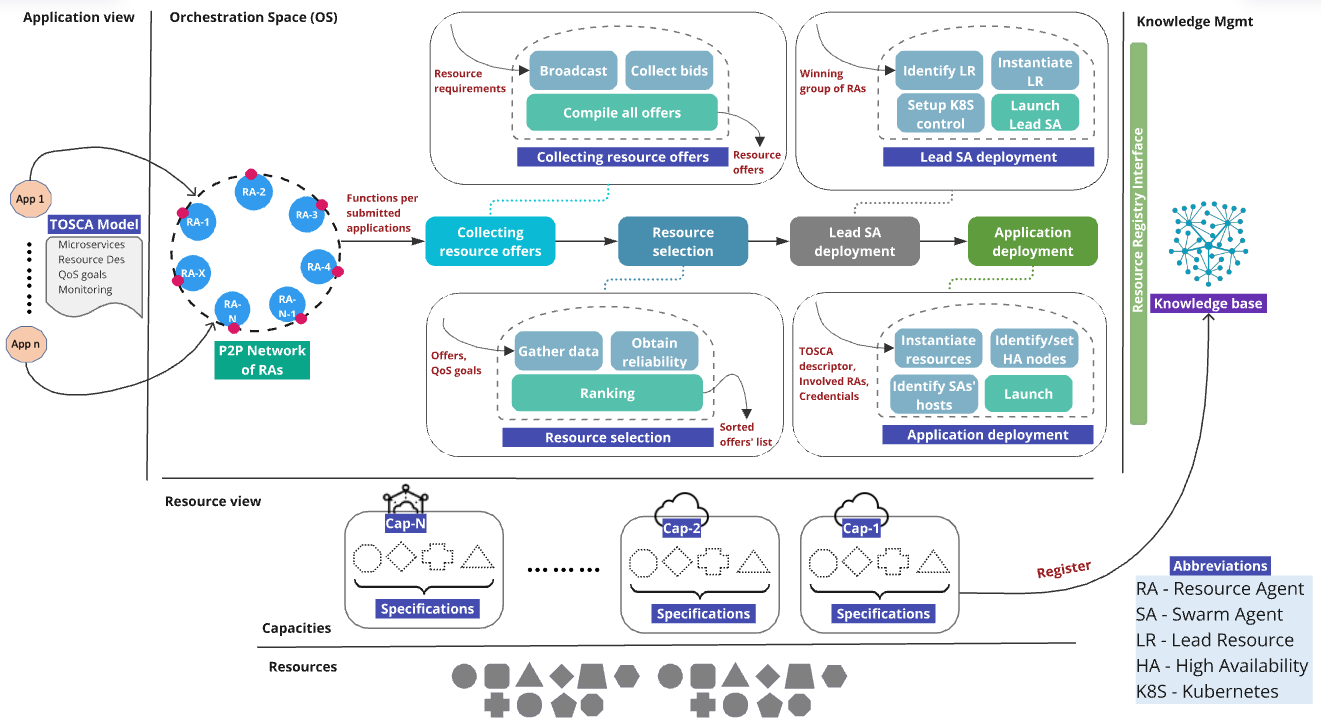}
        \caption{Swarmchestrate architecture}
        \label{fig:framework}
\end{figure*}
This section outlines the high-level Swarmchestrate orchestration concept. Orchestration typically involves the optimised deployment of an application's microservices and any necessary runtime reconfiguration actions. The scope of this paper is limited to the deployment phase; hence, the high-level overview and the accompanying proof of concept refer exclusively to this stage. However, it is important to note that, Swarmchestrate will consider runtime reconfiguration in the next stage.   

Figure~\ref{fig:framework} shows the Swarmchestrate application deployment process, organised into four key sections.
\textbf{Application view} allows operators to define their applications using the standardised TOSCA (Topology and Orchestration Specification for Cloud Applications) application description format~\cite{tosca}. The \textbf{Resource view} comprises a two-layered structure, where \textit{Resources} represents the actual computational resources provided by different cloud and edge providers, e.g. Amazon, Microsoft, and others, and \textit{Capacity}, refers to the logical grouping of resources made available by an entity within the Swarmchestrate ecosystem. In Swarmchestrate, every resource is part of a Capacity, which must be registered in the Knowledge Management system to enable discovery and utilisation for application deployment.

The \textbf{Knowledge Management} component serves as a distributed knowledge base, handling information related to resource capabilities, their descriptions, system interactions, and decision-making processes. Resources can be discovered based on various contextual attributes and trust factors, facilitating efficient application deployment and dynamic reconfiguration. Lastly, the \textbf{Orchestration Space (OS)} is a decentralised entity operating without central control, functioning as the execution environment for core orchestration functions. 

The OS is the confluence of three key concepts---Decentralisation, Swarms, and Intelligence---for achieving efficient, optimised, and trusted application orchestration in the Cloud-Edge continuum. Decentralisation stems from the adoption of a distributed architecture, enabling the system to operate without central control. The Swarm is an adaptation of swarm computing, facilitating the execution and management of individual applications as dynamic, cooperative Swarms. Intelligence is embedded through the extensive use of machine learning and optimisation algorithms, driving resource selection and decision-making for reconfigurations. The following subsections break down the OS components in detail.

\subsection{Application}\label{sec:application}
Swarmchestrate supports microservices-based applications, each consisting of $m$ components. These applications are described using the TOSCA format, which encompasses four key aspects: 
\begin{enumerate}
    \item Component Specification: Details of microservice containers, including requirements like environment variables, commands, and CPU/RAM limits.

    \item Resource Requirements: Specific needs for application resources, such as cloud/edge instances, instance types, hardware limits (CPU/RAM/Storage), and network capabilities.

    \item QoS Goals: Desired QoS specifications, including performance, cost, energy efficiency, trust, service placement, and resource provider/types, to be maintained throughout the application's lifecycle.

    \item Monitoring Specification: Custom metrics to be monitored by Swarmchestrate for informed application-level reconfiguration.
\end{enumerate}
The TOSCA description is submitted to the OS interface, a network of independent Resource Agents (Next section) connected through a peer-to-peer (P2P) system. 
\subsection{Resource Agent}\label{sec:resource-agent}
The Resource Agent (RA), a key system component, is responsible for two main functions. First, it represents one or more Capacities, enabling access to their resources. Second, it facilitates the discovery of suitable resources across the entire resource stack for submitted applications, working collaboratively with other RAs. In Swarmchestrate, an RA is instantiated, when the Capacity provider registers the Capacity resources with attributes like processing power, memory, hardware type, VM instances, pricing, locality, and energy metrics. Once instantiated, the RA connects to other RAs via a P2P network, forming a decentralised OS interface. The application description is submitted to the interface, where an RA in the P2P network receives it and initiates the deployment process, as outlined in the next section.
\begin{figure*}[t]
    \centering
    \begin{subfigure}{0.4\textwidth}
        \centering
        \includegraphics[width=\textwidth]{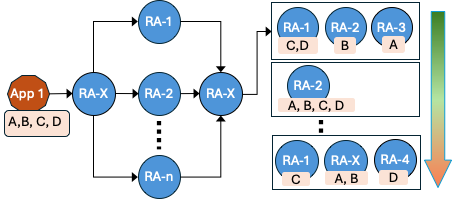}
        \caption{Resource offers collection and ranking}
        \label{fig:offers}
    \end{subfigure}
    \begin{subfigure}{0.3\textwidth}
        \centering
        \includegraphics[width=\textwidth]{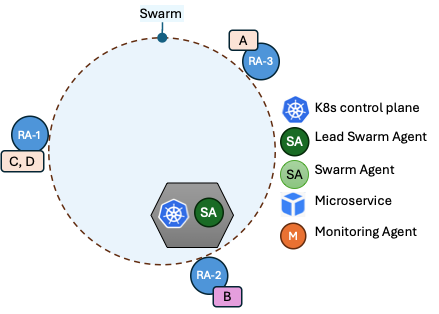}
        \caption{Lead resource}
        \label{fig:leadresource}
    \end{subfigure}
    \begin{subfigure}{0.2\textwidth}
        \centering
        \includegraphics[width=\textwidth]{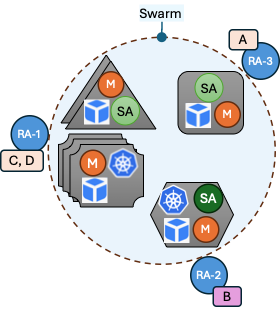}
        \caption{Post-deployment}
        \label{fig:post-deployment}
    \end{subfigure}
    \caption{Illustrative example of application deployment in Swarmchestrate}
\end{figure*}
\subsection{Overall deployment process}\label{sec:overallDeployment}
We illustrate the deployment process using a simple example, featuring an application (app1) comprising four microservices, having four resource requirements (A, B, C, and D). Upon receiving the application (Figure~\ref{fig:offers}), RA-X---randomly selected currently, however can be based on any particular logic---initiates the following steps:
\subsubsection{Collecting resource offers}
The first step is to gather all potential resource sets that meet the application's requirements (A, B, C, and D). As shown in Figure~\ref{fig:offers}, RA-X broadcasts these requirements to all available RAs, requesting offers. Each receiving RA determines its coverage by interacting with the Knowledge Management to assess its Capacity and current usage. Coverage from each RA is classified as Partial (some requirements met), Full (all requirements met), or Zero (no requirements met). RAs respond with their coverage, and RA-X compiles unique groups of possible offers.
\subsubsection{Resource selection}
This step identifies the best offer from all the possible ones obtained in the previous step. Specifically, the best offer represents the optimal set of resources that maximises the chances of fulfilling the QoS goals for the submitted application. Once identified and configured, this set of resources is called Swarm, which will serve the application. To determine the best offer, the following inputs are considered:
\begin{itemize}
    \item The resource offers from the previous step,
    \item The application QoS goals from the TOSCA description (Section~\ref{sec:application}),
    \item The dynamically obtained reliability metrics for each resource offer, represent the overall impact on the achievement of QoS goals. These factors include failure frequency, availability, resource accuracy, performance, network accessibility, and others.
\end{itemize}    
An optimisation algorithm (a candidate implementation is discussed in Section~\ref{sec:ranking}) processes these inputs to generate a ranked list of offers. The first offer on the list is the winner. The resources from the RAs involved in the winning offer (e.g., RA-1, RA-2, and RA-3 in Figure~\ref{fig:offers}) will then be used for the application.
\subsubsection{Lead SA deployment}~\label{sec:leadSA}
This step initiates the formation of the Swarm---the group of resources selected from the best offer---and deploys the lead Swarm Agent (SA). The SA, a system component, is responsible for managing the Swarm. Multiple SAs, launched at the Swarm level, are collectively responsible for keeping the swarm alive and overseeing its self-organisation, including monitoring and making reconfiguration decisions. The first instance of the SA is called the lead SA, responsible for assembling the entire Swarm (discussed in Section~\ref{sec:appDeployment}). The "Lead SA deployment" step begins with the winning group of RAs. If any part of the offer is unavailable, for reasons such as an expired offer or an inaccessible resource, the next-ranked offer is selected. The main task at this stage is to setup a cluster of resources from the involved RAs, which is completed by RA-X using the following sub-steps:
\begin{enumerate}
    \item Identify Lead Resource (LR): RA-X selects the suitable resource from the offer to host the lead SA and Kubernetes---a container orchestration system---control plane. In recent years, Kubernetes has emerged as the de facto standard for container orchestration, establishing itself as the preferred choice over alternative systems such as Docker Swarm and OpenShift. This dominance has informed our decision to adopt Kubernetes for our orchestration framework. Factors such as resource size (CPU, storage), unused capacity, resource type (Cloud or Edge), networking aspects, and rankings from the selection process are considered for the selection of LR. The LR may change dynamically if the resource becomes unavailable or a more suitable one is added during reconfiguration.

    \item Instantiate LR: Once identified, the LR is instantiated (e.g., a cloud resource is dynamically created).

    \item Set up Kubernetes control plane: The Kubernetes control plane is set up on the LR to manage container orchestration for the Swarm.
    
    \item Launch Lead SA: The control plane will initially launch the lead SA, responsible for assembling the entire Swarm (as detailed in the next section).
\end{enumerate}
Completing the above steps results in the state depicted in Figure~\ref{fig:leadresource}. The circle represents the Swarm, comprising four resource types from three RAs. However, at this point, only the LR (type B from RA-2) is live, with the Kubernetes control plane and the lead SA running on it.

\subsubsection{Application deployment}\label{sec:appDeployment}
In this step, the Lead SA completes Swarm formation and application deployment. Inputs include the TOSCA application description, RA details, and required credentials. The task is completed using the following four steps:
\begin{enumerate}
    \item The Lead SA, in collaboration with the involved RAs, instantiates the remaining required resources, which are then connected to the cluster of the LR. 

    \item To prevent a single point of failure, this step identifies additional suitable resources to host a highly available cluster. The selection criteria for the HA node/s are the same as those used for the LR. Once identified, the required HA configurations are implemented.
    
    \item For self-organisation, a group of SAs is responsible. These SAs are hosted on resources alongside the application components. At this step, the Lead SA determines the number of SA instances and their hosting resources within the Swarm.

    \item Using the Kubernetes cluster scheduler, the Lead SA launches the additional SAs, application components, and monitoring agents on the resources. 
\end{enumerate}
The completion of the above steps leads to the status shown in Figure \ref{fig:post-deployment}. At this stage, all required resources---depicted as grayed boxes (two triangles, three plaques, one diamond, and one rectangle)---are part of the same Swarm, running four microservices of app1.

\section{Evaluation}\label{sec:evaluation}
The implementation of the proposed architecture is currently ongoing, therefore, a first study was carried out in a simulation environment to assess feasibility and performance. For this purpose, we used DISSECT-CF-Fog~\cite{markus_2024}, a discrete event simulator popular for its high realism and customisability for Cloud-Edge simulations, to assess the effectiveness of the proposed approach. DISSECT-CF-Fog enables realistic modelling of physical and virtual resources, including their associated energy consumption and network models. This tool was extended with necessary constructs (e.g. RA, Capacity, etc) to support the proposed Swarmchestrate approach. The valuable insights gained from this extended simulation tool will guide the implementation of Swarmchestrate. The implementation and results are open-source and accessible on GitHub\footnote{DISSECT-CF-Fog: https://github.com/sed-inf-u-szeged/DISSECT-CF-Fog}.

\subsection{Application submission and offer selection}

\begin{table}[h]
\caption{Experimental settings for simulation}
	\begin{subtable}{0.2\textwidth}
		\centering
        \scriptsize
		\begin{tabular}{|l|p{15mm}|}
        \hline
		\textbf{Parameter} & \textbf{Value}\\
        \hline \hline
		Num of components  & Compute:3 \newline Storage: 1\\
        \hline
		CPU (pc) & min:1, max:6 \\
        \hline
		RAM (GB) & min:1, max:6 \\
        \hline
		Storage (GB) & min:1, max:10 \\
        \hline
		Image size (MB) & min:1, \newline  max:500 \\
        \hline
		Instances required & min:1, max:3 \\
        \hline
		Message size (KB) & 2 \\
        \hline
		\end{tabular}
		\caption{Applications}
		\label{tab:application}
	\end{subtable}
    \hfill
    \begin{subtable}{0.2\textwidth}
		\centering
        \scriptsize
		\begin{tabular}{|l|l|}
        \hline
		\textbf{Parameter} & \textbf{Value}\\
        \hline \hline
		Location & EU, US\\
		\hline
		Provider & AWS, Azure\\
        \hline
		CPU (pc) & min:16, max:100 \\
        \hline
		RAM (GB) & min:16, max:100 \\
        \hline
		Storage (GB) & min:16, max:100 \\
        \hline
		Idle power (W) & min:150, max:225 \\
        \hline
		Max power (W) & min:500, max:3500 \\
        \hline
		Latency (ms) & min:15, max:100 \\
        \hline
		Price (€/hour) & min:0.025, max:25 \\
        \hline
		\end{tabular}
		\caption{Capacities (Nodes)}
		\label{tab:capacities}
	\end{subtable}
	\label{tab:params}
\end{table}

The simulator receives TOSCA-based application descriptions, as discussed in Section~\ref{sec:application}. The application components can be of two types: \textit{Compute} and \textit{Storage}. A \textit{Compute} component includes a container image for instantiation and hardware limits in terms of CPU and memory. The \textit{Storage} component can only be specified with the size of the allocated partition. Additional aspects such as the provider and the location are optional in both cases. Six applications with different specifications are simultaneously submitted to evaluate the system behaviour in response. Table~\ref{tab:application} presents the ranges used for the specification of each application.

The RA, in simulation, is modelled as a lightweight virtualised resource, utilising 1 CPU core and 1 GB of memory of the underlying physical resource. The physical resources in simulation mimic the behaviour of Capacities (discussed in Section~\ref{sec:resource-agent}). The capacities in simulation environments for our experiments are created from the interval-based specification presented in Table~\ref{tab:capacities}. A capacity with properties closer to the upper bound is considered cloud-based, and closer towards the lower bound is imitating edge-based. The RAs are hosted by these nodes, representing and providing the remaining part of the underlying resource for application components. For our experimentation, the infrastructure consisted of 8 capacities each represented by one RA, jointly forming a network of 8 RAs. 

When an application is submitted, the network randomly assigns an RA to manage the deployment. The assigned RA (RA-X as shown in Figure~\ref{fig:offers}) broadcasts the request to all RAs. Each receiving RA evaluates the request by matching application components to its available capacity using a first-fit strategy, where components are sorted by CPU requirements, with 50\% of RAs sorting it in ascending order and the rest in descending order. The RA maps as many components as its capacity permits. The successful mapping leads to \textit{reserved} state for the concerned subset of capacity. This mixed sorting strategy helps us avoid scenarios, where only components with higher CPU requirements are mapped to capacities. When RA-X receives all responses (including its own)---containing RA-application component pairs---it generates offers by creating all unique combinations of these pairs, ensuring each application component is included exactly once in each combination. These offers are then ranked based on the various characteristics of the involved capacities, employing various optimisation methods detailed in the following section.
%\begin{figure*}[t]
%    \centering
%    \includegraphics[width=\textwidth]{figures/figure_RAs.jpg}
%    \caption{Resource offer submission and acknowledgment of the resource agents for the deployment}
%    \label{fig:distributed_offers}
%\end{figure*}

\subsection{Resource Selection Methods}\label{sec:ranking}
The key purpose of ranking is to select the optimal resource set, which provides the highest chance of fulfilling the QoS objective of submitted applications. For our experiments, the QoS objective of each application consists of 4 key attributes, including latency, cost, bandwidth, and energy consumption. In addition, each attribute is associated with a priority, indicating, the importance of the respective attribute from the application owner's perspective. Furthermore, as explained in Section~\ref{sec:application}, we also consider the reliability of each offer in the decision-making. For ranking, we employed the following two approaches. 
\subsubsection{\textbf{The Cost Function}}
\label{subsection_cost_func}
%\noindent$\rightarrow$ The cost function and how the best-ranked offer is found should be explained. \\
%$\rightarrow$ Ismail will add a simple equation that outlines the cost function along with some explanation here
A cost function was developed to evaluate and rank resource offers based on the employed four QoS attributes. This approach assigns a cost value to each resource offer by normalising and weighting each QoS attribute according to the associated priority. The ranking is then performed based on the overall cost in descending order. More formally, for each QoS attribute $q \in Q$, the raw data $r_q$ is normalised using \eqref{eq:1}.

\begin{align}
\label{eq:1}
\text{\small normalised}_q = \begin{cases} 
\ \ \ \ \ \ \ 0, & \text{\small if \ } \text{\small \textit{max}}(r_q)  = \text{\small \textit{min}}(r_q) \\ 
\frac{r_q - \min(r_q)}{\max(r_q) - \min(r_q)}, & \text{\small otherwise}
\end{cases}
\end{align}

The normalisation ensures that each QoS metric is scaled to a comparable range between 0 and 1. For attributes such as bandwidth, where higher values are preferable, the normalised values are inverted to reflect their desirability. The normalised data is then multiplied by the corresponding QoS priority weight $p_q$. The total cost for an offer $i$ is computed in \eqref{eq:2}.

\begin{align}
\label{eq:2}
\text{total\_cost}_i = \sum_{q \in Q}{p_q \cdot \text{normalised}_{q,i}}
\end{align}

Lastly, to consider the reliability of an offer (denoted by $R$), two variants are employed: 1) \textbf{Additive}, where the reliability factor is subtracted from the total cost (expressed as $\text{total\_cost}_i - R_i$), reducing the cost for more reliable offers; 2) \textbf{Multiplicative}, where the reliability factor scales the total cost (expressed as $\text{total\_cost}_i =  (1-R_i) \cdot \text{total\_cost}_i$).
\subsubsection{\textbf{Borda Voting Method}}
\label{subsection_borda}
%\noindent$\rightarrow$ The voting approach is to be explained here. Ismail is in charge.
The Borda voting, a voting-based ranking algorithm, is also employed to assign scores to each resource offer based on their relative positions across the QoS attributes. In this approach, each QoS attribute is sorted independently based on an assigned preference, e.g., bandwidth in descending order, while latency is in ascending order. The Borda count assigns scores based on rank, with the top-ranked offer receiving the highest score and the tied offers sharing the highest score for their rank. Scores are weighted by attribute priorities and summed to rank offers.

Lastly, similar to the cost function approach, the Borda approach can also incorporate reliability into the ranking process, either as an additive or multiplicative approach. More formally, Equation~\eqref{eq:borda_score} defines the final Borda score $S_i$ of an offer $i$, where $\text{score}_{q}(i)$ and $\text{score}_{R}(i)$ represent the Borda scores for QoS attribute $q$ and reliability $R$, respectively; whereas, Equation \eqref{eq:reliability_additive} and \eqref{eq:reliability_multiplicative} represents the final Borda scores with reliability as additive and multiplicative factors.

\begin{align}
\label{eq:borda_score}
S_i = \sum_{q \in Q}{p_q \cdot \text{score}_{q}(i)}
\end{align} 

\begin{align}
\label{eq:reliability_additive}
S_i = \text{score}_{R}(i) + \sum_{q \in Q}{p_q \cdot \text{score}_{q}(i)}
\end{align}

\begin{align}
\label{eq:reliability_multiplicative}
S_i = R_i \sum_{q \in Q}{p_q \cdot \text{score}_{q}(i)}
\end{align}

\subsection{Application deployment}
The first offer in the ranked list, created through methods discussed in Section~\ref{sec:ranking}, is the winning group consisting of pairs of RA and application components. The next step involves the selection of the lead resource (LR) by RA-X ( explained in Section~\ref{sec:leadSA}). For this experimentation, we consider the capacity with higher CPU cores as the criterion for LR. Furthermore, to mimic realistic behaviour, we also set up a Docker Hub-like image registry to store the container images of the application components with 1000 Mbps bandwidth in the DISSECT-CF-Fog simulator. Once the LR is selected, the image files are transferred over the network where they (the application components) are deployed. Once deployed, the status of associated capacities is changed to \textit{allocated}. Though the primary goal of this paper is to mimic the deployment phase, we also emphasise the long-term impact of the deployment decision by employing a 30-minute task of each deployed \textit{Compute} component to utilise the allocated CPU cores fully. 

\subsection{Results}
\begin{table*}[ht!]
\centering
\caption{Simulation results for different priorities and resource selection methods}
\label{tab:res}
\begin{tabular}{|cc|c|c|c|c|}
\hline
\multicolumn{1}{|c|}{\textbf{Priority}}                                          & \textbf{Method} & \textbf{Simulation Time (min)}          & \textbf{Total Price (EUR)}             & \textbf{Avg. Deployment Time (min)}    & \textbf{Total Energy (KWh)}            \\ \hline\hline
\rowcolor[HTML]{E2E2E2} 
\multicolumn{1}{|c|}{\cellcolor[HTML]{E2E2E2}}                          & Borda  & \cellcolor[HTML]{E6B8AF}37.946 & 0.053                         & 3.636                         & 2.232                         \\ \cline{2-6} 
\rowcolor[HTML]{E2E2E2} 
\multicolumn{1}{|c|}{\multirow{-2}{*}{\cellcolor[HTML]{E2E2E2}Energy}}  & Cost   & \cellcolor[HTML]{E6B8AF}37.946 & 0.046                         & \cellcolor[HTML]{E6B8AF}3.779 & \cellcolor[HTML]{B6D7A8}2.170 \\ \hline
\rowcolor[HTML]{FFFFFF} 
\multicolumn{1}{|c|}{\cellcolor[HTML]{FFFFFF}}                          & Borda  & 35.044                         & \cellcolor[HTML]{B6D7A8}0.015 & 2.451                         & 2.292                         \\ \cline{2-6} 
\rowcolor[HTML]{FFFFFF} 
\multicolumn{1}{|c|}{\multirow{-2}{*}{\cellcolor[HTML]{FFFFFF}Price}}   & Cost   & 37.946                         & 0.032                         & 3.645                         & 2.211                         \\ \hline
\rowcolor[HTML]{E2E2E2} 
\multicolumn{1}{|c|}{\cellcolor[HTML]{E2E2E2}}                          & Borda  & 34.641                         & 0.077                         & 1.352                         & 2.360                         \\ \cline{2-6} 
\rowcolor[HTML]{E2E2E2} 
\multicolumn{1}{|c|}{\multirow{-2}{*}{\cellcolor[HTML]{E2E2E2}Latency}} & Cost   & 34.562                         & 0.079                         & 1.285                         & \cellcolor[HTML]{E6B8AF}2.363 \\ \hline
\multicolumn{1}{|c|}{}                                                  & Borda  & \cellcolor[HTML]{FFFFFF}34.385 & \cellcolor[HTML]{FFFFFF}0.075 & \cellcolor[HTML]{FFFFFF}1.288 & \cellcolor[HTML]{FFFFFF}2.316 \\ \cline{2-6} 
\multicolumn{1}{|c|}{\multirow{-2}{*}{Bandwidth}}                       & Cost   & \cellcolor[HTML]{B6D7A8}32.175 & \cellcolor[HTML]{E6B8AF}0.115 & \cellcolor[HTML]{B6D7A8}0.968 & \cellcolor[HTML]{FFFFFF}2.260 \\ \hline
\rowcolor[HTML]{E2E2E2} 
\multicolumn{1}{|c|}{\cellcolor[HTML]{E2E2E2}}                          & Borda  & 34.318                         & 0.036                         & 1.620                         & 2.229                         \\ \cline{2-6} 
\rowcolor[HTML]{E2E2E2} 
\multicolumn{1}{|c|}{\multirow{-2}{*}{\cellcolor[HTML]{E2E2E2}Equal}}   & Cost   & 34.562                         & 0.076                         & 1.285                         & \cellcolor[HTML]{E6B8AF}2.363 \\ \hline
\multicolumn{2}{|c|}{Random}                                                     & 34.437                         & 0.082                         & 1.365                         & 2.310                         \\ \hline
\end{tabular}
\end{table*}

%\begin{figure*}[ht]
%    \centering
%    \includesvg[width=0.49\textwidth]{figures/energy-prior-chart.svg} \hfill
%    \includesvg[width=0.49\textwidth]{figures/latency-prior-chart.svg}
%    \caption{Energy consumption per node with energy priority (left) and latency priority (right)}\label{fig:res}
%\end{figure*}

\begin{figure*}[ht]
    \centering
    \begin{subfigure}{0.48\linewidth}
        \centering
        \includegraphics[width=\textwidth]{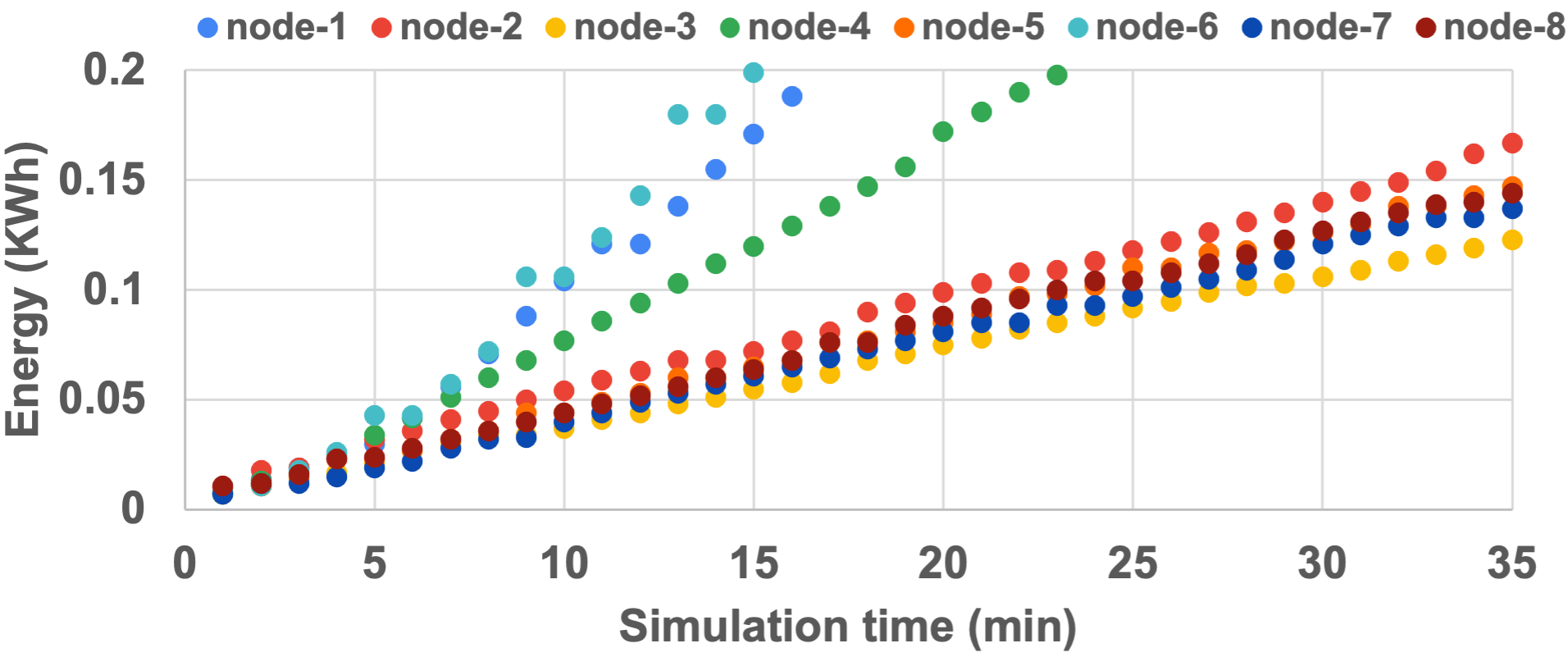}
    \end{subfigure}
    \begin{subfigure}{0.48\linewidth}
        \centering
        \includegraphics[width=\textwidth]{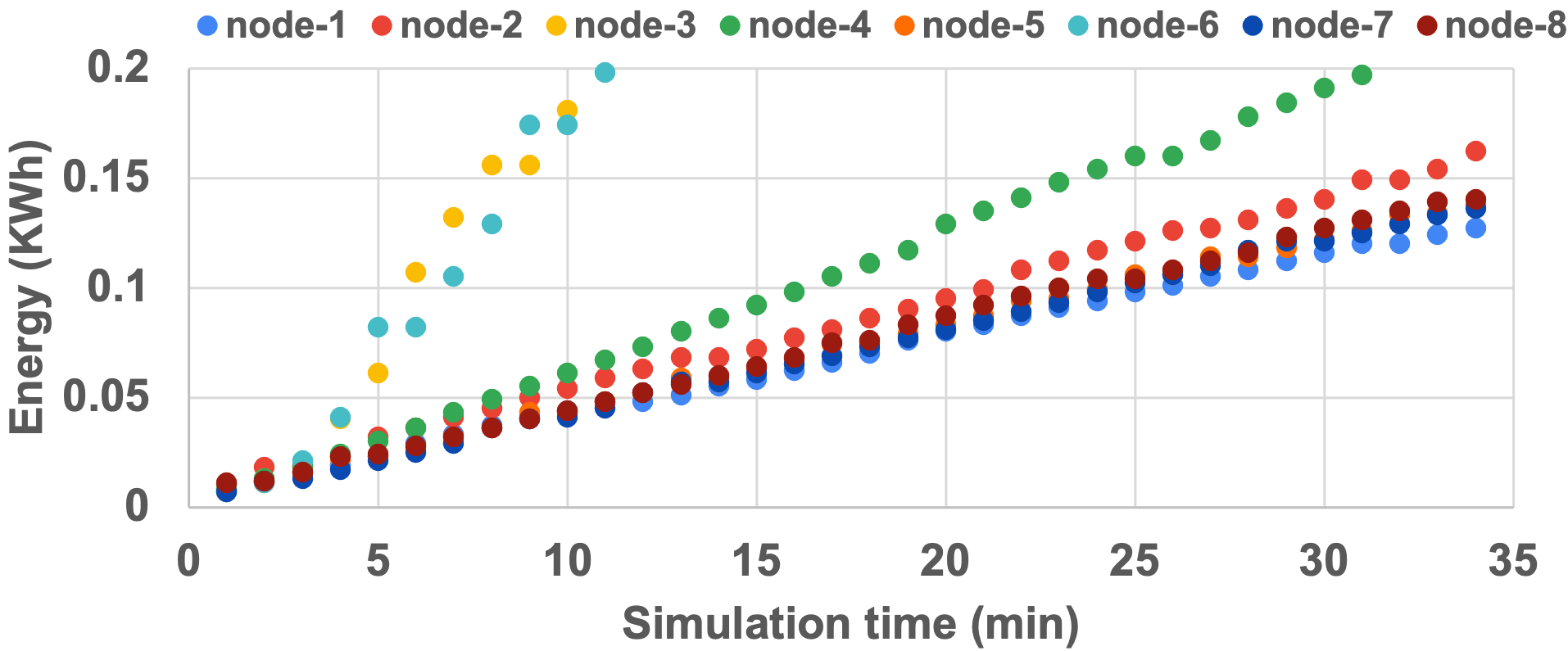}
    \end{subfigure}
    \caption{Energy consumption per node with energy priority (left) and latency priority (right)}
    \label{fig:res}
\end{figure*}

The evaluation examined the ability of Swarmchestrate to handle the simultaneous submission of six applications, with an emphasis on varying priority levels of QoS attributes. We evaluated six strategies: a randomly selected offer, one with equal priorities, and four where a single QoS attribute had a priority of 1.0 while the others were set to 0.1. For comparison, the following metrics are considered: 1) \textit{Simulation Time}, the time from the application submission to the last task's completion; 2) \textit{Total Price}, the resource costs based on hourly provider rates; 3) \textit{Avg. Deployment Time}, the time from submission to component deployment influenced by latency and bandwidth; and 4) \textit{Total Energy}, the aggregated energy consumption per node during the \textit{Simulation Time}.

The results are shown in Table~\ref{tab:res}, where the best values are highlighted in green and the worst in red for each metric, demonstrating the performance of the Borda and Cost approaches. In cases where a priority value of 1.0 was assigned (rows 1-8), the proposed ranking algorithm consistently performed better, as indicated by the green columns. For example, a price-aware strategy can result in the minimisation of operating costs. In terms of approaches, the Cost method, though yields the worst results (red columns), generally outperforms the Borda method in achieving the best priority-specific values. The Equal method balances cost and deployment efficiency, while the bandwidth-aware approach outperforms the latency-aware approach, highlighting its critical role in the deployment process.

More specifically on energy usage, Figure~\ref{fig:res} shows the accumulated energy consumption per node recorded during the energy and latency-aware scenarios. The measurement is for the time from the application submission to the completion of all tasks and does not include the cold start and initial infrastructure setup. In the case of the energy-aware approach, the CPU-heavy tasks start after the fifth minute, while the latency-aware strategy results in faster application deployment (as it is shown by \textit{Avg. Deployment Time} in Table \ref{tab:res} as well), thus the related tasks already start after the third minute.
\section{Conclusion}\label{sec:conclusion}
This study introduced Swarmchestrate, a decentralised orchestration framework for managing applications in the Cloud-Edge continuum. Swarmchestrate leverages an application-centric approach to address key challenges such as scalability, resource heterogeneity, self-organisation and the need to balance multiple QoS objectives. Through simulation-based evaluation, we demonstrated the potential and applicability of the proposed framework in terms of seamless application deployment across diverse resource providers. 
Building on these findings, work is currently ongoing to implement the Swarmchestrate concepts and finalise the mechanism of self-organisation for the runtime reconfiguration. The implemented framework will also be prototyped on four real-life industry use cases. These advancements aim to establish Swarmchestrate as a robust and scalable solution for next-generation distributed systems.

\section*{Acknowledgment}
This work was funded by the European Commission's Horizon programme within the Swarmchestrate project, (project no. 101135012).

\bibliographystyle{unsrt} % Sorts references in order of citation
\bibliography{main} % Points to your BibTeX file

%\begin{thebibliography}{00}

%\iffalse
%\bibitem{b1} G. Eason, B. Noble, and I. N. Sneddon, ``On certain integrals of Lipschitz-Hankel type involving products of Bessel functions,'' Phil. Trans. Roy. Soc. London, vol. A247, pp. 529--551, April 1955.
%\bibitem{b2} J. Clerk Maxwell, A Treatise on Electricity and Magnetism, 3rd ed., vol. 2. Oxford: Clarendon, 1892, pp.68--73.
%\bibitem{b3} I. S. Jacobs and C. P. Bean, ``Fine particles, thin films and exchange anisotropy,'' in Magnetism, vol. III, G. T. Rado and H. Suhl, Eds. New York: Academic, 1963, pp. 271--350.
%\bibitem{b4} K. Elissa, ``Title of paper if known,'' unpublished.
%\bibitem{b5} R. Nicole, ``Title of paper with only first word capitalized,'' J. Name Stand. Abbrev., in press.
%\bibitem{b6} Y. Yorozu, M. Hirano, K. Oka, and Y. Tagawa, ``Electron spectroscopy studies on magneto-optical media and plastic substrate interface,'' IEEE Transl. J. Magn. Japan, vol. 2, pp. 740--741, August 1987 [Digests 9th Annual Conf. Magnetics Japan, p. 301, 1982].
%\bibitem{b7} M. Young, The Technical Writer's Handbook. Mill Valley, CA: University Science, 1989.
%\fi

%\bibitem{markus_2024} A. Markus, V. D. Hegedus, J. D. Dombi and A. Kertesz, `Fuzzy-based Task Offloading with Machine Learning-driven Forecasting for IoT`, 2024 IEEE 8th International Conference on Fog and Edge Computing (ICFEC), pp. 71-78, 2024. DOI: 10.1109/ICFEC61590.2024.00015
%\end{thebibliography}
\vspace{12pt}

\end{document}